\documentclass[12pt,a4paper]{article}

\usepackage[utf8]{inputenc}

\usepackage{graphicx}
\usepackage{hyperref}
\usepackage[margin=2cm]{geometry}
\usepackage{times}
\usepackage[textsize=footnotesize]{todonotes}
\usepackage{pifont}
\usepackage{amssymb}
\usepackage{amsmath}
\usepackage{units}
\usepackage{algorithm} 
\usepackage{algpseudocode} 
\usepackage{setspace}

\usepackage[style=apa,natbib]{biblatex}
\title{A note on efficient audit sample selection}
\author{Laura Boeschoten, Sander Scholtus \& Arnout van Delden}
\date{}

\begin{document}
\maketitle

\begin{abstract}
    Auditing is a widely used method for quality improvement, and many guidelines are available advising on how to draw samples for auditing. However, researchers or auditors sometimes find themselves in situations that are not straightforward and the standard sampling techniques are not sufficient, for example when a selective sample has initially been audited and the auditor desires to re-use as many cases as possible from this initial audit in a new audit sample that is representative with respect to some background characteristics. In this paper, we introduce a method that selects an audit sample that re-uses initially audited cases by considering the selection of a representative audit sample as a constrained minimization problem. In addition, we evaluate the performance of this method by means of a simulation study and we apply the method to draw an audit sample of establishments to evaluate the quality of an establishment registry used to produce statistics on energy consumption per type of economic activity.
    
    \bigskip
    
\end{abstract}

\section{Introduction}
{\it Auditing} is a widely used method for quality improvement. The goal of an audit is to estimate the extent to which practice in a field of profession complies with certain identified review criteria. The degree of compliance helps in identifying areas where improvements can be made \citep{hearnshaw2003audits}. As practical reasons often limit the ability to apply an audit to the total population, typically a sample is drawn on which the audit is performed, the \emph{audit sample}. The procedure of performing an audit is known under different denominators. An audit can be performed in the sense that a sample is thoroughly evaluated, or a high quality test can be applied to the audit (which is then often known as the \emph{gold standard}). The study of comparing (potentially) error-prone outcomes to true scores is also known as a \emph{validation study} and the set of \emph{true scores} is then known as a \emph{validation sample}. For supervised machine learning a manually annotated set is created also referred to as a \emph{labelled set}. Audits are performed in different fields of profession.

In the field of clinical research, audits can for example be used to evaluate the performance of diagnostics tests, as applied by \citet{chataway2004herpes}. In such cases, an audit is used to evaluate the performance of a certain test that is used to diagnose a specific type of disease. This typically means that a diagnostic test is available and of high quality (`gold standard'). However, it is often expensive, difficult to apply or invasive and therefore has a high patient burden. An alternative, cheaper, more easy to use, lower patient burden diagnostic test is also available, although it has a lower quality in the sense that more outcomes are either false positives or false negatives compared to the gold standard test. To evaluate the quality of the new diagnostic test, both tests are applied to the audit sample and the outcomes of the new test are compared to outcomes of the `gold standard' test.

In many organizations, financial audits are performed, sometimes because it is required by law \citep{World2019}. Here, the extent to which an organization's financial statements are presented fairly is evaluated \citep{derks2019jasp}. Within an organization, all financial transactions can be considered a population, from which a sample is drawn, sometimes by means of a risk function, from which the total amount of financial misstatements in the organization is inferred \citep{elder2013audit}.

In official statistics, audits are also being used to estimate the quality of administrative sources or of surveys. This can for example be done simultaneously by using a structural equation model as shown by \citet{scholtus2015error} including an audit sample \citep{sobel1986platonic}. 
Another field where the use of an audit sample plays a role is in supervised machine learning. There, a manually annotated set with (error-free) labels is used to train and test a machine learning model. The test set is used to estimate the performance of the model for the intended population. Therefore, for the test set it is important that it is representative for the target population. When one is interested in predicting a class with a small proportion in the population, then using a simple random sample to obtain an annotated set is not so efficient. One may for instance use a set of key words to create a selective sample. The question then is how, starting from a part of the selective sample, it can be supplemented to obtain a representative test set. In the field of machine learning, this issue is also known as `sample selection bias' \citep{zadrozny2004learning} and is for example manifested when a road network is inferred from Weigh-in-Motion road sensors \citep{klingwort2021using}.

As an audit is often too expensive and/or invasive to be performed on the complete population, typically a sample is selected on which the audit is performed. Ideally, this sample is selected at random from the complete population directly or from a Randomized Controlled Trial (RCT). In such cases, it can be assumed that the audit sample is representative for the population. In practice however, auditors often face situations that are more complex. Examples are when an observational study is performed instead of an RCT, when the auditor prefers to draw a judgement sample or when the register that is available as a sampling frame is incomplete. In such cases, the sample should be selected more carefully in order for it to be representative. Representativity can for example be achieved by investigating background characteristics of the sample units and by drawing a stratified sample. Typically, guidelines are also available for drawing stratified samples, such as provided by the British Medical Journal \citep{ougrin2006clinical} or the American Institute of Certified Public Accountants \citep{bell1995auditing}.

In practice however, auditors find themselves often in situations that are more complex and for which the standard guidelines are not sufficient. A first example of such a situation is when the auditor has determined a maximum number of units for which auditing is feasible or when the auditor has limitations in terms of budget. A second example of a complex situation is when a small non-random sub-sample of the population has already been audited which should ideally be included in the audit sample or when the auditor plans to include particular units in the audit sample with certainty. In such situations, using standard guidelines for auditing are not sufficient.

In this paper, we propose to approach the selection of an audit sample as a constrained minimization problem. In the proposed framework, a representative audit sample can be straightforwardly selected for the non-standard situations described above. In Section 2, notation and the proposed framework are introduced. In Section 3, it is illustrated how the framework can be used for the complex audit situations. In Section 4, a simulation study is used to investigate the performance of the method under different conditions. Section 5 illustrates how the proposed method can be applied in practice using ready-made R scripts. Finally, section 6 concludes.

\section{Background}
\subsection{A representative sample} \label{represent}
When an audit is performed, the goal is to select a sample that is representative with respect to the target population. We assume that this audit sample is drawn from a subset of units in the target population for which a possibly error-prone measure of the outcome variable of interest is available. We will refer to this subset as the \emph{observed sample}. In some applications, the observed sample is in fact equal to the target population. In general, the observed sample is a sample from the target population, and standard sampling theory applies to both the audit sample with respect to the observed sample and the observed sample with respect to the target population.

When an audit is performed for a diagnostic test in health care, the observed sample refers to the sample on which an inferior test is applied. When an algorithm is validated, the observed sample refers to the set on which the prediction was performed. In official statistics, the observed sample can refer to an administrative data set that is used to compile statistics; in such cases the observed sample and target population are often assumed to be the same. For all above scenarios it holds that both the observed sample and the audit sample should be representative with respect to the population to which inference is desired. 

In what follows, all variables are assumed to be categorical. The observed sample is denoted as $U$, with $U = {1, 2, \ldots, N}$. The true variable of interest is denoted by $W$, with $W_g$ denoting the value of this variable for element $g$ in $U$. The aim is to estimate certain parameters of the distribution of $W$, for instance, its frequency distribution in the target population. It is assumed that $W$ is observed \emph{only} for units that are audited. For all units in the observed sample $U$, an error-prone version of the variable of interest is measured, which is denoted by $X$. The values of $X$ in the observed sample are denoted by $X_1, X_2, \ldots, X_N$. In addition to estimating the distribution of $W$, a secondary aim may be to estimate the association between $W$ and $X$, e.g., the error probabilities $\text{Pr}(X_g = x \: \vert \: W_g = w)$. Knowledge of these error probabilities makes it possible to estimate the distribution of $W$ based on the whole observed sample, not just the audit sample, which is more efficient. In addition, these error probabilities may be of interest in their own right, as quality measures of the error-prone indicator in $X$.

In addition to $X$, we suppose that one or more background variables are available in the observed sample, collectively denoted as $Y$, with values $Y_1,Y_2,\ldots,Y_N$. We assume that the joint distribution of $Y$ is known (or previously estimated) for the target population. The content of the variables in $Y$ depends on the specific audit study of interest. In case of a diagnostic test in healthcare, $Y$ can contain variables such as age, gender, ethnic background and prior health status. In case of a financial audit, $Y$ can contain variables such as employer ID, monetary size or department in firm.

Let $Z$ denote the selection indicator of the audit sample within $U$, where $Z_g$ is the value for element $g$ with $g = 1, 2, \ldots, N$. $Z_g = 1$ when the element is included in the sample and $Z_g = 0$ otherwise. The expected value of $Z_g$ with respect to the sample design is denoted by 
\begin{equation}
    \pi_g = E(Z_g) = \text{Pr}(Z_g = 1).
\end{equation}
The audit sample size, denoted by $n$, is equal to the sum of the values of $Z$:
\begin{equation}
    n = \sum_{g=1}^{N} Z_g.
\end{equation}

In order to make appropriate inferences regarding the target population, it is important that the audit sample is representative with respect to the target population. If the observed sample was randomly drawn from that population and is representative, the audit sample can also be randomly selected from the observed sample by giving every unit an equal inclusion probability ($\pi_g = n / N$).

In practice, situations may occur in which the inclusion probabilities should be allowed to depend on $Y$, for example because some type of elements are more likely or are allowed to decline for audit, because a judgement sample was selected, or because the observed sample was not representative with respect to the target population, for example because some groups of elements were so small that they were over-sampled in the observed sample. In such (in practice very likely) cases a simple random sample without replacement for the audit sample will not be sufficient to obtain representativity.

We will now give a more precise definition of representativity as it will be used in this paper. As a theoretical starting point to investigate the selectivity of an audit sample, the joint distribution in the observed sample of the variables $(W,X,Y,Z)$ is of interest, where $W$ and $X$ represent the true variable of interest and its error-prone observed version, $Z$ indicates whether a unit is included in the audit sample and $Y$ represents a grouping variable based on background characteristics. We make the following simplifying assumption:

\bigskip

\noindent \textbf{Assumption A.} There is no \emph{direct} association between the true variable of interest $W$ and the audit inclusion indicator $Z$, once $X$ and $Y$ are accounted for.

\bigskip

\noindent For any audit sample that we draw ourselves, we can ensure that this assumption is satisfied. Alternatively, if we are given a  possibly non-random sample of previously audited units, the assumption may require that the right background variables are included in $Y$. The importance of this assumption which will be seen below is that it allows the selectivity of the audit sample to be investigated by analysing the joint distribution of $(X,Y,Z)$ instead of $(W,X,Y,Z)$ (see the end of Section~\ref{dev}). In practice, the former distribution is known, whereas inference about the latter distribution becomes possible only once a representative audit sample has been obtained.

Selectivity in whether or not a unit is included in the audit sample (\emph{audit inclusion}),  meaning that the distribution of $Z$ depends on $X$, is not problematic as long as differences in inclusion probabilities are completely related to the background characteristics in $Y$ for which the distribution in the target population is known:
\begin{equation} \label{eq:MAR}
\text{Pr}(Z_g = 1 \: \vert \: X_g = x, Y_g = y) = \text{Pr}(Z_g = 1 \: \vert \: Y_g = y).
\end{equation}
From a missing data perspective, it can be said that audit inclusion is not problematic as long as exclusion is Missing At Random (MAR) \citep{rubin1976inference}. However, if the distribution of $Z$ depends on $X$ and this dependence is not explained by $Y$, this can be problematic, in particular when the error probabilities $\text{Pr}(X_g = x \: \vert \: W_g = w)$ are of interest. From a missing data perspective, it can be said that exclusion is then Missing Not At Random (MNAR) \citep{rubin1976inference}. Therefore, our goal is to select an audit sample in such a way that \eqref{eq:MAR} holds; we refer to this as the \emph{MAR selectivity requirement}.

\subsection{Deviance as a criterion for representativity} \label{dev}

To determine whether a given audit sample meets the MAR selectivity requirement \eqref{eq:MAR}, we propose to analyse the joint distribution $(X,Y,Z)$ in the observed sample using a non-saturated log-linear model $(XY)(YZ)$. Note that the difference between the non-saturated model and a saturated model is that the direct association $(XZ)$ and the three-way interaction $(XYZ)$ are excluded. In the non-saturated model it is assumed that $X$ and $Z$ are conditionally independent given $Y$, i.e. the observed outcome variable and the audit inclusion probability are conditionally independent given the background characteristics under consideration. Therefore, we refer to it as the \emph{independence model}.

The number of observations in cell $(X=i, Y=j, Z=k)$ is denoted as $n_{ijk}$ and the number of observations estimated by the independence model is denoted as $\hat{n}_{ijk}$. The fit of the independence model can be measured by the likelihood ratio test statistic or deviance ($D$) comparing it to the saturated model where $\hat{n}_{ijk} = n_{ijk}$ \citep{agresti2013categorical}:
\begin{align}
    &D = 2 \sum_{i,j,k} n_{ijk} \textrm{ log } n_{ijk} - 2 \sum_{i,j,k} {n}_{ijk} \textrm{ log } \hat{n}_{ijk},
\end{align}
where $D \geq 0$. A higher value for $D$ indicates a stronger conditional dependence between $X$ and $Z$ given $Y$ in the observed data-set, which means that the selectivity problem is more substantive. This suggests that an audit sample should be selected for which $D$ is `sufficiently small'.

For the independence model, the estimated value for $\hat{n}_{ijk}$ can be calculated directly, without the use of an iterative algorithm, as
\begin{equation} \label{explicit}
    \hat{n}_{ijk} = \frac{n_{ij+}n_{+jk}}{n_{+j+}},
\end{equation}
see for example \citet{agresti2013categorical} or \citet{bishopfienbergholland}. Here, $n_{ij+} = \sum_{k} n_{ijk}$, $n_{+jk} = \sum_{i} n_{ijk}$, and $n_{+j+} = \sum_{i,k} n_{ijk}$.

The definition of $D$ can be expressed alternatively by incorporating \eqref{explicit}: 
\begin{align} \label{dev_explicit}
D &= 2 \sum_{i,j,k} n_{ijk} \textrm{ log } n_{ijk} - 2 \sum_{i,j} n_{ij+} \textrm{ log } n_{ij+} \\ \nonumber
  &\quad - 2 \sum_{j,k} n_{+jk} \textrm{ log } n_{+jk} + 2 \sum_{j} n_{+j+} \textrm{ log } n_{+j+}\\ \nonumber
 &= C + 2 \sum_{i,j,k} n_{ijk} \textrm{ log } n_{ijk} - 2 \sum_{j,k} n_{+jk} \textrm{ log } n_{+jk},
\end{align}
where
\begin{equation} \label{C}
C = 2 \sum_{j} n_{+j+} \textrm{ log } n_{+j+} - 2 \sum_{i,j} n_{ij+} \textrm{ log } n_{ij+}
\end{equation}
is a constant term which depends only on the distribution of $(X,Y)$ and therefore will be the same for every possible choice of audit sample. For the second term in the first line of Equation \eqref{dev_explicit} we used that $\sum_{i,j,k} n_{ijk} \textrm{ log } n_{ij+} = \sum_{i,j} n_{ij+} \textrm{ log } n_{ij+}$; for the third and fourth term alike.

Recall from Section~\ref{represent} that the full (unobserved) distribution of interest is $(W,X,Y,Z)$. We will now show that under Assumption A (there is no direct association between $W$ and $Z$), the selectivity of the audit sample can be analysed using the deviance defined in Equation \eqref{dev_explicit}, which is based on the observed distribution $(X,Y,Z)$.

Let $n'_{hijk}$ denote the number of observations in cell $(W = h, X = i, Y = j, Z = k)$, with $n_{ijk} = n'_{+ijk} = \sum_{h} n'_{hijk}$. Under Assumption A, the maximal hierarchical log linear model for $n'_{hijk}$ is $(WXY)(XYZ)$. For this model, the predicted values $\hat{n}'_{hijk}$ can also be obtained directly:
\begin{displaymath}
    \hat{n}'_{hijk} = \frac{n'_{hij+}n'_{+ijk}}{n'_{+ij+}} = \frac{n'_{hij+}n_{ijk}}{n_{ij+}},
\end{displaymath}
see \citet{bishopfienbergholland}. 

Hence, the deviance that compares this model to the saturated model with $\hat{n}'_{hijk} = n'_{hijk}$ can be written analogously to Equation \eqref{dev_explicit} as
\begin{align*}
D_{W,max} &= 2 \sum_{h,i,j,k} n'_{hijk} \textrm{ log } n'_{hijk} - 2 \sum_{h,i,j} n'_{hij+} \textrm{ log } n'_{hij+} \\
  &\quad - 2 \sum_{i,j,k} n_{ijk} \textrm{ log } n_{ijk} + 2 \sum_{i,j} n_{ij+} \textrm{ log } n_{ij+}.
\end{align*}

The maximal model contains a direct association between $X$ and $Z$, pointing to selectivity in the audit sample. The analogue of the independence model for $(W,X,Y,Z)$ is given by the log linear model $(WXY)(YZ)$. For this model, the predicted values satisfy:
\begin{displaymath}
    \hat{n}'_{hijk} = \frac{n'_{hij+}n'_{++jk}}{n'_{++j+}} = \frac{n'_{hij+}n_{+jk}}{n_{+j+}},
\end{displaymath}
see \citet{bishopfienbergholland}.

Hence, the deviance that compares this independence model to the saturated model for $(W,X,Y,Z)$ can be written as
\begin{align*}
D_{W} &= 2 \sum_{h,i,j,k} n'_{hijk} \textrm{ log } n'_{hijk} - 2 \sum_{h,i,j} n'_{hij+} \textrm{ log } n'_{hij+} \\
  &\quad - 2 \sum_{j,k} n_{+jk} \textrm{ log } n_{+jk} + 2 \sum_{j} n_{+j+} \textrm{ log } n_{+j+}.
\end{align*}
However, under Assumption A the saturated model is too large, and it makes more sense to compare the fit of the independence model to that of the above maximal model. This can be done using the conditional deviance for nested models \citep{bishopfienbergholland}, which in this case is given by:
\begin{align*}
D_{W} - D_{W,max} &= 2 \sum_{i,j,k} n_{ijk} \textrm{ log } n_{ijk} - 2 \sum_{i,j} n_{ij+} \textrm{ log } n_{ij+} \\
  &\quad - 2 \sum_{j,k} n_{+jk} \textrm{ log } n_{+jk} + 2 \sum_{j} n_{+j+} \textrm{ log } n_{+j+}.
\end{align*}

Clearly, $D_{W} - D_{W,max} = D$ in Equation \eqref{dev_explicit}. Hence, under assumption A, to analyse the fit of the independence model compared to the maximal model for the full distribution $(W,X,Y,Z)$, it suffices to analyse the fit of the model $(XY)(YZ)$ for the observed distribution $(X,Y,Z)$ .

\section{Method}
In this section, we introduce how a representative audit sample can be selected by using the deviance defined in Equation \eqref{dev_explicit} as a criterion for representativity. The proposed method has two main advantages. First, the auditor can determine a maximum number of units to be audited. Here, the proposed method will provide an audit sample that is as representative as possible given that maximum number. Second, if some units have already been audited or the auditor is planning to include specific units in the audit sample, the method attempts to include as many of these units into the final audit sample as possible, given the constraint that this final sample should be sufficiently representative.

\subsection{Basic optimization problem}
Let $n_{ijk}$ denote the number of observations in cell $(X=i,Y=j,Z=k)$ prior to applying the method in this section. If previously audited units are available, these have $Z = 1$ and so it holds that $n_{ij1} > 0$ for certain units. If the selection of the audit sample starts with a `blank canvas', then $n_{ij1} = 0$ and $n_{ij0} = n_{ij+}$ for all $i$ and $j$. We consider the general situation where additional units may be selected for auditing (moved from $Z = 0$ to $Z = 1$) and previously selected units may be excluded (moved from $Z = 1$ to $Z = 0$). When applying the method, the user should specify a maximum number of additional units to include in the audit sample ($M_{+}$) and a maximum number of previously audited units to exclude ($M_{-}$). Special cases are obtained by setting one of these bounds to zero. Moreover, in the special case of no previously audited units, it automatically holds that $M_{-} = 0$ and $M_{+}$ indicates the maximal size of the audit sample.

After applying the method, the adjusted number of units in cell $(X=i,Y=j,Z=k)$ is denoted as $m_{ijk}$. We write
\begin{equation} \label{mij1}
    m_{ij1} = n_{ij1} + \delta^{+}_{ij} - \delta^{-}_{ij},
\end{equation}
where $\delta^{+}_{ij}$ and $\delta^{-}_{ij}$ indicate the number of additional units to audit and the number of previously audited units to exclude with $(X=i,Y=j)$, respectively. Note that during this procedure, values in the marginal table $(X,Y)$ are not adjusted, only units are transported from $Z=0$ to $Z=1$ and vice versa. Hence, $m_{ij+} = n_{ij+}$ for all $i$ and $j$.

Now, the goal is to sample the (additional) units from different cells of table $(X,Y,Z)$ in such a way that $D$ defined in Equation \eqref{dev_explicit} is minimized. This is a minimization problem for which the target function can be written as: 
\begin{equation}
D(m) = C + 2 \sum_{i,j,k} m_{ijk} \textrm{ log } m_{ijk} - 2 \sum_{j,k} m_{+jk} \textrm{ log } m_{+jk},
\end{equation}
since $C$ remains equal to \eqref{C}, as $m_{ij+} = n_{ij+}$ and $m_{+j+} = n_{+j+}$. Therefore, $C$ can be ignored when minimizing $D(m)$.

When minimizing $D(m)$, a number of constraints apply. First, the user-specified bounds on the number of additional units to include ($M_{+}$) and the number of units to exclude ($M_{-}$) lead to the following constraints: 
\begin{align}
    \sum_{i,j} \delta^{+}_{ij} &\leq M_{+} \\
    \sum_{i,j} \delta^{-}_{ij} &\leq M_{-}.
\end{align}
Second, for each combination $(X=i,Y=j)$, the number of units in- and excluded in the audit sample should add up to the known marginal value: 
\begin{align}
    m_{ij1} + m_{ij0} = n_{ij+},\qquad & i = 1,\dots, I; \\ \nonumber
                                       & j = 1,\dots, J,
\end{align}
where $I$ is the number of categories in $X$ and $J$ is the number of categories in $Y$. Together with $m_{ij1}$ defined in Equation \eqref{mij1}, this restriction implies that
\begin{equation} \label{mij0}
    m_{ij0} = n_{ij0} - \delta^{+}_{ij} + \delta^{-}_{ij}.
\end{equation}

Third, for each combination $(X=i,Y=j)$, there exist bounds on $\delta_{ij}^{+}$ and $\delta_{ij}^{-}$ based on the initial counts $n_{ijk}$, since no more units can be moved from $Z=0$ to $Z=1$ or vice versa than are initially available:
\begin{align}
    0 \leq \delta_{ij}^{+} \leq n_{ij0}, \qquad & i = 1,\dots, I;
    & j = 1 ,\dots, J. \\
    0 \leq \delta_{ij}^{-} \leq n_{ij1}, \qquad & i = 1,\dots, I;
    & j = 1 ,\dots, J.
\end{align}
In combination with the other restrictions, these bounds imply that $m_{ijk} \geq 0$ for every value.

The minimization procedure can now be written as follows: 
\begin{equation} \label{optim}
    \min \Big\{ C + 2 \sum_{i,j,k} m_{ijk} \textrm{ log } m_{ijk} - 2  \sum_{j,k} m_{+jk} \textrm{ log } m_{+jk} \Big\}
\end{equation}
under constraints 
\begin{align*}
    & m_{ij1} = n_{ij1} + \delta^{+}_{ij} - \delta^{-}_{ij}; \\
    & m_{ij0} = n_{ij0} - \delta^{+}_{ij} + \delta^{-}_{ij}; \\
    & \sum_{i,j} \delta^{+}_{ij} \leq M_{+}; \\
    & \sum_{i,j} \delta^{-}_{ij} \leq M_{-}; \\
    & 0 \leq \delta_{ij}^{+} \leq n_{ij0}; \\ 
    & 0 \leq \delta_{ij}^{-} \leq n_{ij1}.
\end{align*}

This is a non-linear minimization problem with linear constraints which can be solved by standard software, for instance using the R function `constrOptim' \citep{R2019}. To find the optimal value efficiently, it is useful to have the gradient of the objective function $D(m)$, i.e., the vector of partial derivatives with respect to the free parameters $\delta_{ij}^{+}$ and $\delta_{ij}^{-}$. For the elements of this gradient, we find [using \eqref{mij1} and \eqref{mij0} and some standard calculus]:
\begin{align}
    \frac{\partial D(m)}{\partial \delta_{ij}^{+}} &= 2 \bigg(\log m_{ij1} - \log m_{ij0} - \log m_{+j1} + \log m_{+j0} \bigg); \label{grad_plus} \\
    \frac{\partial D(m)}{\partial \delta_{ij}^{-}} &= 2 \bigg(\log m_{ij0} - \log m_{ij1} - \log m_{+j0} + \log m_{+j1} \bigg). \label{grad_min}
\end{align}

\subsection{A procedure for optimizing the audit sample} \label{sec:procedure}
In practice, it turns out that the objective function defined in Equation \eqref{optim} is difficult to minimize, because it has many local minima. To ensure that a global minimum is found, multiple starting values have to be tried in the optimization algorithm.

\begin{algorithm}
\caption{Optimize the audit sample \label{algoritm_audit}}
\textbf{input:} The number of observations $n_{ijk}$ ($i=1,\ldots,I; \ j=1,\ldots,J; \ k = 0,1$). \
\begin{algorithmic}[1]
\State Initialize $D_{\text{best}}$ (the best deviance found so far) as the value of $D$ for the original counts $n_{ijk}$.
\State Determine the bounds $M_{+}$ and $M_{-}$.
\State Determine $N_{\text{attempts}}$, the maximal number of attempts to search for the best solution.

\For{1 : $N_{\text{attempts}}$}
        \State Select random starting values for the solution:
        \begin{itemize}
            \item Draw a value $\widetilde{M}_{+}$ from a uniform distribution on $[l_{+}M_{+}, u_{+}M_{+}]$, where $l_{+}$ and $u_{+}$ are chosen constants with $0 < l_{+} < u_{+} < 1$. 
            \item Assign random starting values $\delta_{ij}^{+}$ inside their feasible intervals such that $\sum_{i,j} \delta_{ij}^{+} = \widetilde{M}_{+}$.
            \item Draw a value $\widetilde{M}_{-}$ from a uniform distribution on $[l_{-}M_{-}, u_{-}M_{-}]$, where $l_{-}$ and $u_{-}$ are chosen constants with $0 < l_{-} < u_{-} < 1$. 
            \item Assign random starting values $\delta_{ij}^{-}$ inside their feasible intervals such that $\sum_{i,j} \delta_{ij}^{-} = \widetilde{M}_{-}$.
        \end{itemize}
        \State Run the optimization algorithm to solve \eqref{optim} with these starting values. 
        \State Compute $D(m)$ for the current solution. If $D(m) < D_{\text{best}}$, then set $D_{\text{best}} := D(m)$ and store the current solution. Otherwise, discard the current solution.
\EndFor
\State If $D_{\text{best}}$ is still considered too large, return to step 3 and change the bounds $M_{+}$ and/or $M_{-}$.
\end{algorithmic}
\textbf{output:} The number of additional units to audit $\delta_{ij}^{+}$ and remove from the audit sample $\delta_{ij}^{-}$ ($i=1,\ldots,I; \ j=1,\ldots,J$) found for the best solution, as well as the value of $D_{\mathrm{best}}$.
\end{algorithm}

For solving Equation \eqref{optim}, we propose the practical procedure found under Algorithm~\ref{algoritm_audit}. To help decide when the best deviance value $D_{\text{best}}$ should be considered `too large' in step 9, it may be noted that under the independence model $(XY)(YZ)$ the deviance asymptotically follows a chi-square distribution with $J \times (I-1)$ degrees of freedom \citep{agresti2013categorical}. Hence, the $(1-\alpha) \times 100\%$ percentile $\chi_{J(I-1)}^{2}(1-\alpha)$ of this distribution (e.g., with $\alpha = .05$) could be used as a cut-off point: we accept the current solution in step 9 when $D_{\text{best}} \leq \chi_{J(I-1)}^{2}(1-\alpha)$.

Once the optimal values $\delta_{ij}^{+}$ and $\delta_{ij}^{-}$ have been obtained, a representative audit sample can be obtained by applying the following two steps (in parallel) to each stratum $(X=i,Y=j)$:
\begin{itemize}
    \item If $\delta_{ij}^{+} > 0$, draw a simple random sample without replacement of size $\delta_{ij}^{+}$ from the $n_{ij0}$ units in this stratum with $Z = 0$. These units are selected for additional auditing, so moved to $Z = 1$.
    \item If $\delta_{ij}^{-} > 0$, draw a simple random sample without replacement of size $\delta_{ij}^{-}$ from the (original) $n_{ij1}$ units in this stratum with $Z = 1$. These previously audited units are removed from the audit sample, so moved to $Z = 0$.
\end{itemize}
Once these steps have been run, the final audit sample consists of all units with $Z = 1$.

\subsection{Minimizing the deviance while maximizing the number of re-used cases} \label{sec:recycle}

The approach introduced in Section 3.2 uses deviance as a primary criterion to select an audit sample, where the sample is selected with the lowest value for deviance. Recycling units that have been in the initial audit sample is considered as a secondary criterion. However, in practice a deviance that is as low as possible is not always essential, and recycling as many cases as possible can be more important. Depending on sample size and number of categories in $Y$ a critical value for deviance can be determined and all selected samples with a deviance lower than the critical value can be considered as being representative with respect to $Y$.

More formally stated, the specification of the optimization problem in \eqref{optim} allows that the optimal solution contains combinations $(i,j)$ where $\delta_{ij}^{+} > 0$ and $\delta_{ij}^{-} > 0$ simultaneously. This corresponds to a solution where in the same stratum some units are added to the audit sample while other, previously audited units, are removed from the audit sample. This seems inefficient from a practical point of view.

It is easy to correct this after the solution has been found, by moving to an equivalent solution with
\begin{align*}
    \delta_{ij,alt}^{+} &= \max \left\{ 0, \delta_{ij}^{+} - \delta_{ij}^{-} \right\}, \\
    \delta_{ij,alt}^{-} &= \max \left\{ 0, \delta_{ij}^{-} - \delta_{ij}^{+} \right\}.
\end{align*}
Clearly, $\delta_{ij,alt}^{+} - \delta_{ij,alt}^{-} = \delta_{ij}^{+} - \delta_{ij}^{-}$, so this corresponds to the same solution in terms of $m_{ijk}$, but now it holds that $\delta_{ij,alt}^{+} \delta_{ij,alt}^{-} = 0$ in all cases, as desired. However, it would be preferable to avoid this type of inefficient solution altogether.

A possible solution is to replace the objective function $D(m)$ of the minimization problem defined in Equation \eqref{optim} by one of the following alternatives:
\begin{align}
F_{1}(m) &= D(m) + \lambda \sum_{i,j} \big(\delta_{ij}^{+} + \delta_{ij}^{-}\big), \label{optim_alt1} \\
F_{2}(m) &= D(m) + \exp \left\{ - \frac{D(m)}{\kappa} \right\} \sum_{i,j} \big(\delta_{ij}^{+} + \delta_{ij}^{-}\big). \label{optim_alt2}
\end{align}
where $\lambda$ is a small positive constant (e.g., $\lambda = 0.01$) and $\kappa$ is a positive constant such that $D(m)/\kappa$ is larger than, say, $10$ for any candidate solution with a deviance value for which the independence model would be rejected (as discussed in Section~\ref{sec:procedure}).

Both alternative objective functions penalize candidate solutions with large total numbers of additional units to audit and previously audited units to remove (i.e., large values of $\sum_{i,j} \big(\delta_{ij}^{+} + \delta_{ij}^{-}\big)$). This should make any candidate solution with $\delta_{ij}^{+} \delta_{ij}^{-} > 0$ unattractive, because the value of $F_1(m)$ or $F_2(m)$ can be reduced by replacing $(\delta_{ij}^{+},\delta_{ij}^{-})$ by $(\delta_{ij,alt}^{+},\delta_{ij,alt}^{-})$ as above. The exponential term in $F_{2}(m)$ is supposed to ensure that the penalty term becomes relevant only for candidate solutions with small (i.e., acceptable) values of $D(m)$. For large values of $D(m)$, it holds that $F_{2}(m) \approx D(m)$. The same can be achieved with $F_{1}(m)$, provided that the constant $\lambda$ is chosen with some care.

The problem of minimizing \eqref{optim} with the target function replaced by \eqref{optim_alt1} or \eqref{optim_alt2} can be solved by the same procedure as outlined in Section~\ref{sec:procedure}. To obtain the gradient values for $F_{1}(m)$, we simply add a term $\lambda$ to \eqref{grad_plus} and \eqref{grad_min}. For $F_{2}(m)$ we obtain:
\begin{displaymath}
 \frac{\partial F_{2}(m)}{\partial x} = \frac{\partial D(m)}{\partial x} + \exp \left\{ - \frac{D(m)}{\kappa} \right\} \left\{ 1 - \frac{1}{\kappa} \frac{\partial D(m)}{\partial x} \sum_{i,j} \big(\delta_{ij}^{+} + \delta_{ij}^{-}\big) \right\},
\end{displaymath}
where $x = \delta_{ij}^{+}$ or $x = \delta_{ij}^{-}$.

\section{Simulation study}
\subsection{Simulation approach for deviance and bias} \label{sec:approach}
To empirically evaluate the performance of the audit sample selection procedure, we conducted a simulation study using R \citep{R2019}. The code used for the simulation study is available on \url{https://github.com/lauraboeschoten/audit_2021} and a small illustrative example of the code applied to one situation is available at \url{https://github.com/lauraboeschoten/audit_2021/tree/master/reproducible_example}.

The main simulation study consists of three sets of conditions. For each set, theoretical populations with varying relationships between $W$, $X$, $Y$ and $Z$ are generated, where $X$ and $Y$ have three categories and $Z$ has two categories by design. Note that in this simulation study, the target population and observed sample are the same. The first set investigates how the performance of the proposed procedure might be affected by the strength of the relationship between $X$ and $Z$ before the start of the optimization procedure, varying from no relationship to a strong relationship. The second set investigates how the performance might be affected by the strength of the relationship between $W$ and $X$, varying from a perfect relationship (meaning that observed variable $X$ is a perfect measurement of the outcomes after audit $W$) to an imperfect and asymmetrical relationship (meaning that observed variable $X$ contains measurement error and that the probability of an incorrect score differs for different scores of the audit variable $W$). The third set investigates how the performance of the procedure might be affected by the strength of the relationship between $W$ and $Y$, varying from a strong relationship to a weak relationship. Note that for each set, the conditions are ordered from most desired situation to least desired situation. Each set contains four conditions resulting in a total of twelve simulation conditions, which can all be found in Table \ref{tab:simconds}. Note that when the different relationships between $X$ and $Z$ are investigated, the first conditions listed in Table \ref{tab:simconds} for $(W,X)$ and $(W,Y)$ are selected and similar approaches are used when investigating $(W,X)$ and $(W,Y)$ to investigate the main effects of these relationships. In the last part of the study, the interactions between the selected relations are investigated by taking the most desired and least desired condition for each relation, and investigating their combinations in a full factorial design. 

In all conditions, a joint probability density is generated by multiplying the probabilities listed in Table \ref{tab:simconds}. The probabilities generated here follow the log-linear model $(WX)(WY)(XZ)$. This model is contained within the previously discussed  maximal model $(WXY)(XYZ)$ and it contains a term $(XZ)$ that cannot be found in the independence model $(WXY)(YZ)$. Because of this term, it is expected that this model results in a large value for the deviance and it would therefore be beneficial to adjust the audit sample via the proposed minimization procedure. For each of the described conditions, $1,000$ data-sets of size $N = 10,000$ are sampled from the generated joint probability density. The size of the audit sample is $300$ in all conditions, as follows also from the proportions described in Table \ref{tab:simconds}.

\begin{table}[h!]
\centering
\begin{tabular}{ccccccccccccccccc} 
\hline \hline 
 \multicolumn{6}{c}{Simulation condition number} \\
        & &   & 1 &   & &   & 2 & & &   & 3 & & &   & 4 &\\
 \cline{3-5} \cline{7-9} \cline{11-13} \cline{15-17}
        & & X &   &   & & X &   &   & & X &   &   & & X &   &   \\
        & & 1 & 2 & 3 & & 1 & 2 & 3 & & 1 & 2 & 3 & & 1 & 2 & 3 \\ \hline
 Z & 0  & .323 & .323 & .323 & & 
          .323 & .323 & .323 & &
          .323 & .323 & .323 & &
          .323 & .323 & .323 \\
   & 1  & .010 & .010 & .010 & &  
          .012 & .010 & .008 & &
          .015 & .010 & .005 & &
          .018 & .010 & .002  \\\\
        & & W &   &   & & W &   &   & & W &   &   & & W &   &   \\
        & & 1 & 2 & 3 & & 1 & 2 & 3 & & 1 & 2 & 3 & & 1 & 2 & 3 \\\hline   
 X & 1  & .333 & 0    & 0    & & 
          .267 & .033 & .033 & & 
          .333 & 0    & 0    & & 
          .300 & .017 & .017 \\
   & 2  & 0    & .333 & 0    & & 
          .033 & .267 & .033 & & 
          .017 & .300 & .017 & & 
          .033 & .267 & .033 \\
   & 3  & 0    & 0    & .333 & & 
          .033 & .033 & .267 & & 
          .033 & .033 & .267 & & 
          .050 & .050 & .233 \\ \\
Y  & 1  & .267 & .033 & .033 & & 
          .200 & .067 & .067 & &
          .300 & .017 & .017 & & 
          .267 & .033 & .033 \\
   & 2 &  .033 & .267 & .033 & & 
          .067 & .200 & .067 & & 
          .033 & .267 & .033 & & 
          .067 & .200 & .067 \\
   & 3 &  .033 & .033 & .267 & &
          .067 & .067 & .267 & &
          .050 & .050 & .233 & &
          .100 & .100 & .133 \\ \hline\hline
\end{tabular}
\caption{Overview of the bivariate relationships used to generate data for the simulation study}
\label{tab:simconds}
\end{table}

Per condition, the procedure as described in Section \ref{sec:procedure} is applied on each generated data-set. The bounds $M_{+}$ and $M_{-}$ that should be determined at step 2 of Algorithm~\ref{algoritm_audit} are set at $100$ and $10$ respectively and the maximal number of attempts for the optimization procedure to search for a solution at step 3 is set at $N_{\text{attempts}} = 200$. Note that step 9 (adjusting the bounds $M_{+}$ and $M_{-}$ if the obtained value for the deviance is considered too large) is left out to appropriately compare the results under different conditions. The settings described here remain consistent for all simulation conditions. 

We considered two types of target parameters to estimate in this simulation study: the true proportion of units in the population with $W_g = w$ for each category $w$,
\begin{equation} \label{eq:PW}
P^{W}_{w} = \frac{1}{N} \sum_{g=1}^{N} I(W_g = w),
\end{equation}
and the proportion of units in the population with true category $W_g = w$ that are observed in category $X_g = x$ (measurement error probabilities):
\begin{equation} \label{eq:PX|W}
    P^{X|W}_{x|w} = \frac{\sum_{g=1}^{N} I(W_g = w, X_g = x)}{\sum_{g=1}^{N} I(W_g = w)}.
\end{equation}
Under the assumption that the independence model holds, $P^{W}_{w}$ defined in \eqref{eq:PW} can be estimated without bias from the audit sample by
\begin{equation} \label{eq:est_PW}
    \hat{P}^{W}_{w} = \sum_{y} P^{Y}_{y} \frac{\sum_{g=1}^{n} I(W_g = w, Y_g = y)}{\sum_{g=1}^{n} I(Y_g = y)} \equiv \sum_{y} P^{Y}_{y} p^{W|Y}_{w|y},
\end{equation}
where $P^{Y}_{y}$ is the (known) proportion of units with $Y_g = y$ in the population and $p^{W|Y}_{w|y}$ denotes the (unweighted) observed proportion of cases with $Y_g = y$ in the audit sample that also have $W_g = w$. (Here, for convenience it is assumed that the audit sample consists of the first $n$ units.) Similarly, $P^{X|W}_{x|w}$ defined in \eqref{eq:PX|W} can be estimated consistently from the audit sample by:
\begin{equation} \label{eq:est_PX|W}
    \hat{P}^{X|W}_{x|w} = \frac{\sum_{y} P^{Y}_{y} \frac{\sum_{g=1}^{n} I(W_g = w, X_g = x, Y_g = y)}{\sum_{g=1}^{n} I(Y_g = y)}}{\sum_{y} P^{Y}_{y} \frac{\sum_{g=1}^{n} I(W_g = w, Y_g = y)}{\sum_{g=1}^{n} I(Y_g = y)}} \equiv \frac{\sum_y P^{Y}_{y} p^{WX|Y}_{w,x|y}}{\sum_y P^{Y}_{y} p^{W|Y}_{w|y}}.
\end{equation}
Note that this is a special case of the so-called combined ratio estimator \citep{cochran}.

We investigate the bias of ${P}^{W}_{w}$,  $B({P}^{W}_{w}) = E(\hat{P}^{W}_{w} - P^{W}_{w})$ and of $\hat{P}^{X|W}_{x|w}$, $B(\hat{P}^{X|W}_{x|w}) = E(\hat{P}^{X|W}_{x|w} - P^{X|W}_{x|w}$) while selecting only $Z=1$ both before and after the procedure and compare these results.

\subsection{Simulation approach for variance} \label{sec:simvar}
In practice, it may be relevant to estimate the (design-based) variance of $\hat{P}^{W}_{w}$ and $\hat{P}^{X|W}_{x|w}$ from the observed data. Under the assumption that the independence model holds, we propose to estimate these variances by treating the audit sample as a stratified simple random sample with $Y$ as a stratifying variable; see the remark at the end of this subsection for a justification. For $\hat{P}^{W}_{w}$ this yields the following variance estimator, assuming for simplicity that the sampling fraction in each stratum is small enough so that finite population corrections can be neglected:
\begin{equation} \label{eq:var_PW}
    \widehat{\mathrm{var}}(\hat{P}^{W}_{w}) = \sum_{y} \frac{(P^{Y}_{y})^2}{n_y} p^{W|Y}_{w|y} (1 - p^{W|Y}_{w|y}),
\end{equation}
where $n_y$ denotes the number of units with $Y_g = y$ in the audit sample. Similarly, under the same assumptions the approximate variance of $\hat{P}^{X|W}_{x|w}$ can be derived from expression (6.51) in \citet{cochran}. 
Assuming that $N P^{Y}_y \gg 1$ for all $y$, we obtain the following variance estimator:
\begin{align}
    \widehat{\mathrm{var}}(\hat{P}^{X|W}_{x|w}) &= \frac{1}{(\hat{P}^{W}_{w})^2} \sum_{y} \frac{(P^{Y}_{y})^2}{n_y} \left\{ p^{WX|Y}_{w,x|y} (1 - p^{WX|Y}_{w,x|y}) + (\hat{P}^{X|W}_{x|w})^2 p^{W|Y}_{w|y} (1 - p^{W|Y}_{w|y}) \right. \nonumber \\
    &\phantom{=} \qquad \left. - 2 \hat{P}^{X|W}_{x|w} p^{WX|Y}_{w,x|y} (1 - p^{W|Y}_{w|y}) \right\}. \label{eq:var_PX|W}
\end{align}

Note that these design-based variances treat the target population of size $N$ as fixed. To evaluate the performance of these variance estimators, a separate set of simulations was run. Combining the least desired condition of $(W,X)$ in Table~\ref{tab:simconds} (i.e., the right-most condition) with each of the four conditions of $(W,Y)$, four fixed target populations of size $N = 10,000$ were generated. From each of these populations, $1,000$ initial audit samples were generated according to the least desired condition of $(X,Z)$ in Table~\ref{tab:simconds}. Again, the procedure from Section~\ref{sec:procedure} was applied to each audit sample, with the same settings as before. For each final audit sample, we computed $\hat{P}^{W}_{w}$ and $\hat{P}^{X|W}_{x|w}$ as well as the associated variance estimates from \eqref{eq:var_PW} and \eqref{eq:var_PX|W}. This allowed us to compare the estimated variances with the empirical variances of $\hat{P}^{W}_{w}$ and $\hat{P}^{X|W}_{x|w}$ across $1,000$ simulation rounds.

\emph{Remark:} To justify the assumption used to derive \eqref{eq:var_PW} and \eqref{eq:var_PX|W} --- that the final audit sample resembles a stratified simple random sample --- we note that the independence model $(WXY)(YZ)$ is equivalent to a logistic regression model for $Z$ that, besides the constant term, uses only $Y$ as a predictor; cf.~\citet[Section 9.5.1]{agresti2013categorical}. Hence, the distribution of the sample size $n_y$ per stratum does not depend on the target parameters $P^{W}_{w}$ and $P^{X|W}_{x|w}$: the sample sizes $n_y$ are \emph{ancillary statistics} with respect to these target parameters. As discussed by \citet{holtsmith1979}, it is preferable to evaluate the variance of estimated parameters conditional on any ancillary statistics. Hence, in our case the variances of $\hat{P}^{W}_{w}$ and $\hat{P}^{X|W}_{x|w}$ should be evaluated conditional on the realised sample sizes $n_y$. Moreover, if the independence model holds after the optimization procedure, according to the above interpretation as a logistic regression model, all units in the same stratum based on $Y$ have the same (final) inclusion probability. Therefore, when the realised sample sizes $n_y$ are treated as fixed, in the absence of more information it makes sense to consider the design of the audit sample as equivalent to a stratified simple random sample.

\subsection{Results}

\subsubsection{Deviance}

\begin{figure}[h!]
\caption{Results in terms of relative deviance (see text). Boxplots demonstrate the spread of the 1,000 relative deviance values obtained per simulation condition. The three panels represent the three sets of bivariate relationships specified in the study, and the four columns per panel illustrate the four alternative strengths of those relationships specified (see Table 1). \label{fig:deviance}}
\centering
\includegraphics[width=\textwidth]{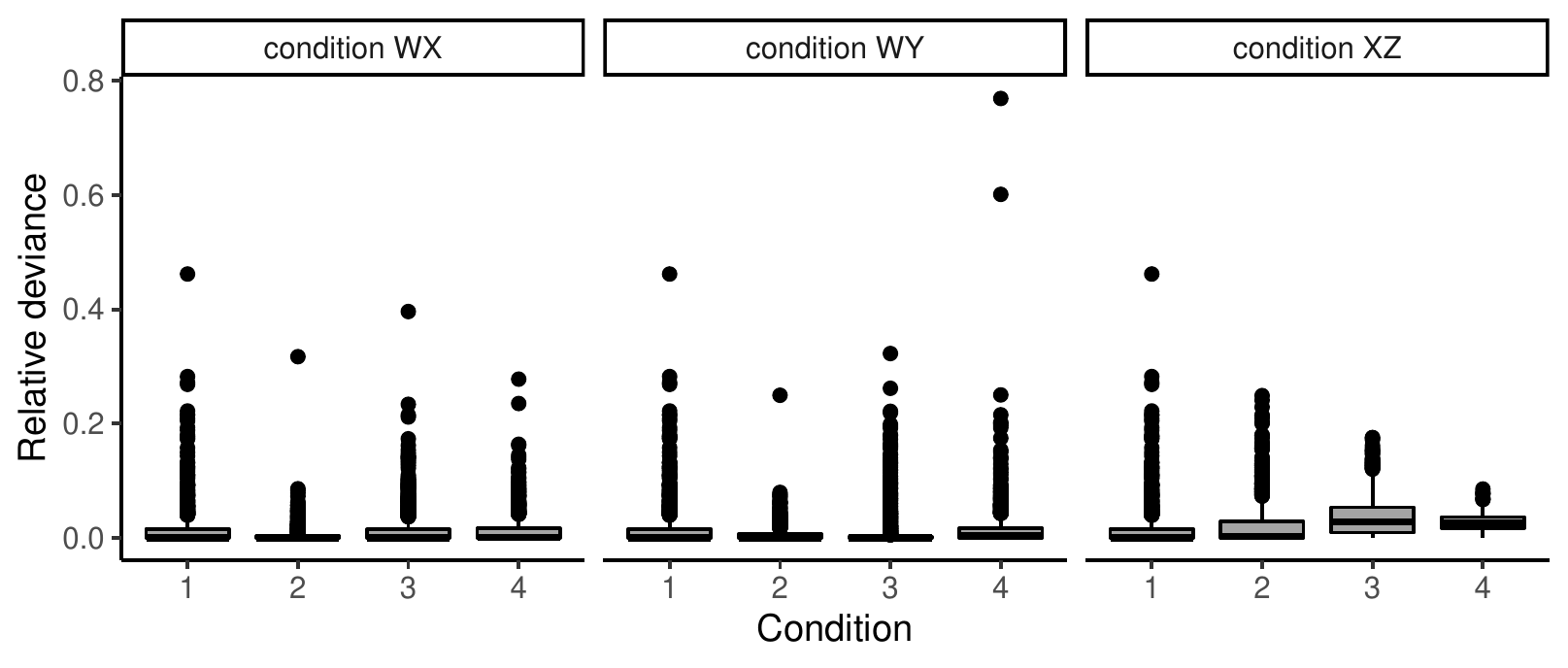}
\end{figure}

Figure \ref{fig:deviance} illustrates the different values obtained for relative deviance under the different simulation conditions. By relative deviance we mean the deviance after applying the optimization procedure as a proportion of the deviance before applying the procedure, so a value close to zero means that a new audit sample is drawn that substantially improves the representativity of the audit. 

In the left panel, the results are shown for the four different $WX$ relationships under which the initial sample was drawn. As representativity is defined with respect to $Y$, not with respect to $W$, differences in relative deviance between these $WX$ conditions are not expected and this is confirmed here. In the middle panel, the results are shown for the four different $WY$ relationships under which the initial sample was drawn. The (lack of) differences between these boxplots illustrates that the proposed method is able to perform in situations where the relationship between background variables and the variable of interest in the audit sample is strong and possibly also unbalanced. In the right panel, results are shown for the four different $XZ$ relationships under which the initial audit sample was drawn. Here, a small increase in average relative deviance can be detected when inclusion in the initial audit sample relates more strongly to scores on $X$, while the spread of the results becomes drastically smaller. This is likely to be caused by the fact that the deviance of the initial model was more substantive. 

\begin{figure}[h!]
\caption{Results in terms of relative deviance (see text). Boxplots demonstrate the spread of the 1,000 relative deviance values obtained per simulation condition for eight different combinations of different  strengths of relationships between the variables $(WX)$, $(WY)$ and $(YZ)$ (see Table 1). \label{fig:deviance_ff}}
\centering
\includegraphics[width=\textwidth]{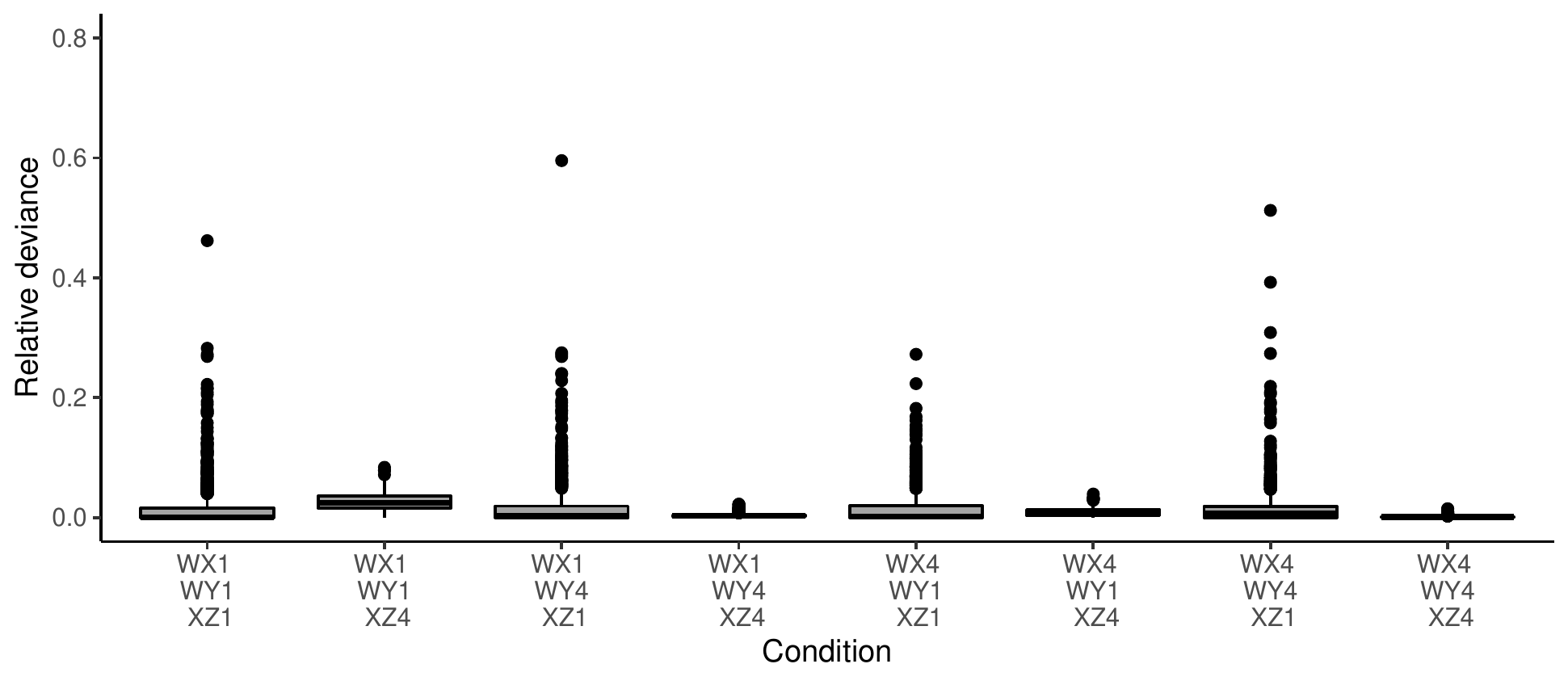}
\end{figure}

Figure \ref{fig:deviance_ff} illustrates the different values obtained for relative deviance under interactions between the different simulation conditions. As indicated by the boxplot labels, for each boxplot, the starting dataset is generated under a different combination of conditions, for which the complete overview can be found in Table 1. These boxplots illustrate  that the scores for relative deviance are stable over different data-generating conditions. It is particularly noteworthy that a strong relationship for $(XZ)$ apparently has the most substantial influence on scores for relative deviance. 

\subsubsection{Bias in W}

\begin{figure}[h!]
\caption{Boxplots of the bias distribution of the estimated proportions of $W$, based on 1,000 replicates per simulation condition. The three panels represent the three sets of bivariate relationships specified in the study, and the four columns per panel illustrate the four strengths of those relationships (see Table 1). \label{fig:biasW}}
\centering
\includegraphics[width=\textwidth]{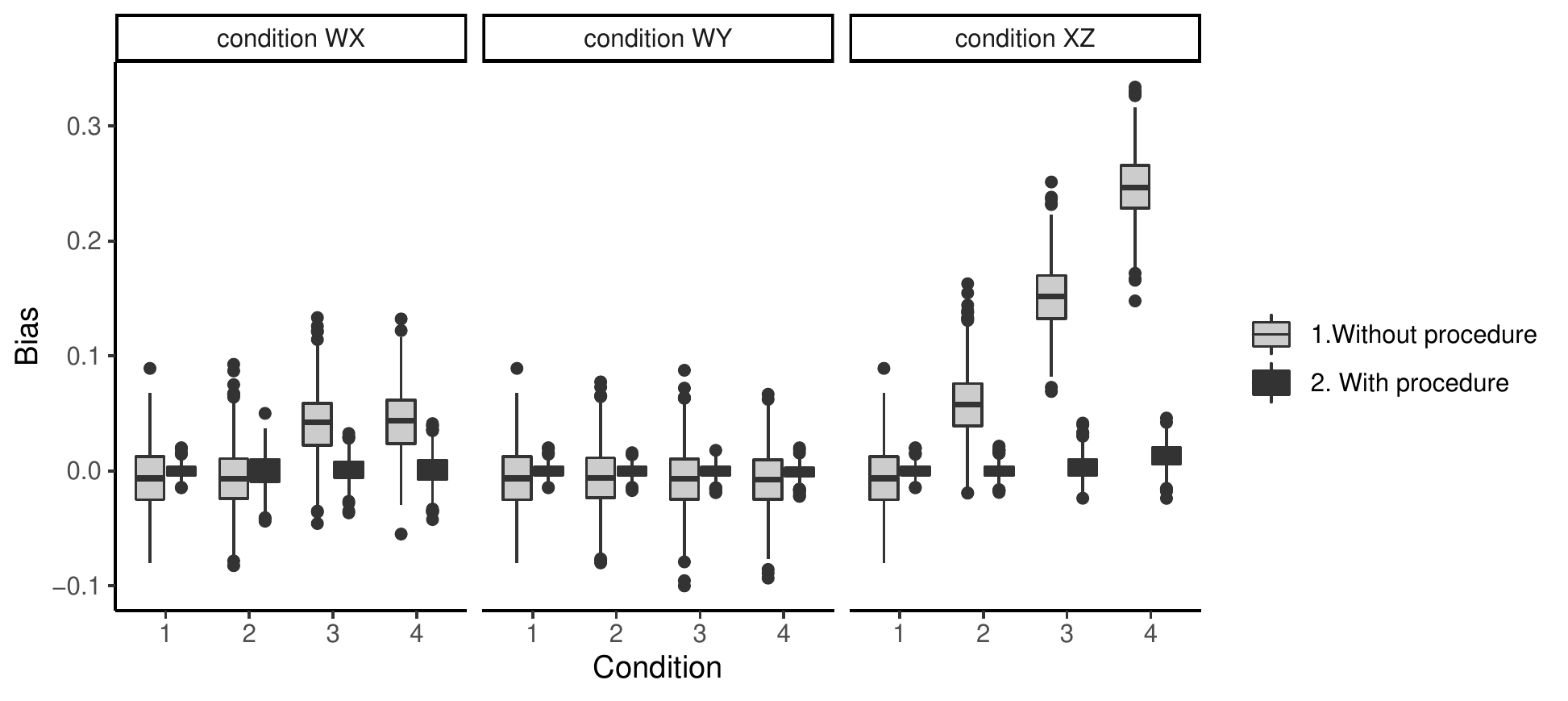}
\end{figure}

Figure \ref{fig:biasW} illustrates the different values obtained for bias under the different simulation conditions compared to when the initial audit sample would have been used directly. Only the results for $W=1$ are shown as the results for $W=2$ and $W=3$ behaved very similarly.

In the left panel, the results are shown for the four different $WX$ relationships under which the initial sample was drawn. In the first condition $W=X$. As $X$ is the observed target variable of interest and $W$ its true version, it is expected that no bias is present in results for $W$ if $W=X$. In conditions 2-4, $W\neq X$ and small increases in spread can be detected as a consequence. However, it can also be seen that if the procedure is not applied in such cases, bias is induced in the results for $W$. In the middle panel, the results are shown for the four different $WY$ conditions. Similar to the results found for the deviance, it can be concluded that the relationship between the variable of interest and background variables does not affect the bias in $W$. However, bias is also not detected before the procedure was applied, so a correction procedure is not required in such situations. In the right panel, results are shown for the four different $XZ$ conditions. Here, a small increase in bias can be detected when inclusion in the initial audit sample relates more strongly to scores on $X$, i.e. the procedure finds it more difficult to obtain a sample that is unbiased with respect to $W$ if inclusion in the initial sample becomes more unbalanced with respect to $X$. However, not applying the procedure in such situation results in very substantive amounts of bias in $W$. 

\begin{figure}[h!]
\caption{Boxplots of the distribution of the bias of the estimated proportions of $W$, based on 1,000 replicates per simulation condition for eight combinations of different strengths of relationships between the variables $(WX)$, $(WY)$ and $(YZ)$ (see Table 1). \label{fig:biasW_ff}}
\centering
\includegraphics[width=\textwidth]{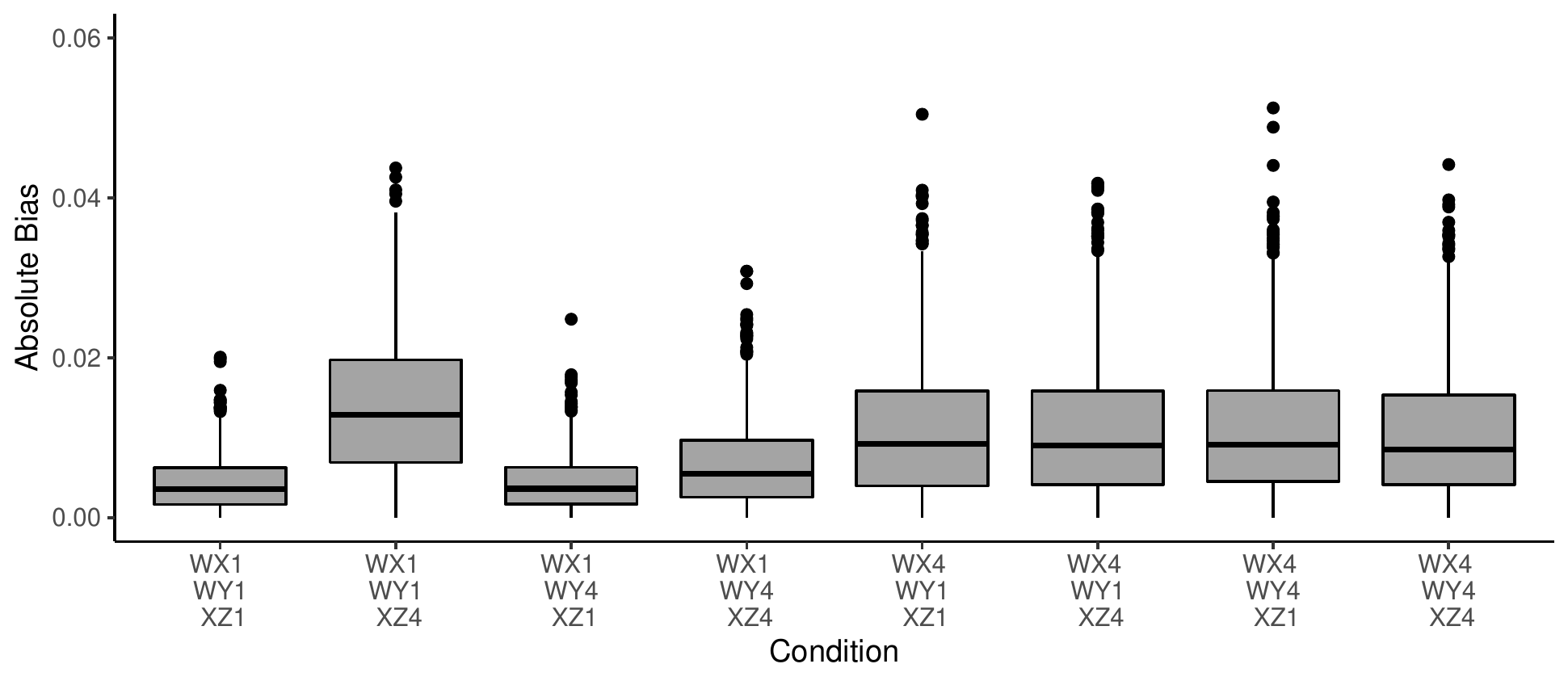}
\end{figure}

Figure \ref{fig:biasW_ff} illustrates the different values of bias in $W$ after applying the method under interactions between the different simulation conditions. As indicated by the boxplot labels, for each boxplot, the data is generated under a different combination of conditions. These boxplots illustrate that although the bias present in estimates of $W$ can be caused by both a weaker relationship in $(WX)$ or a stronger relationship in $(XZ)$, the combination of these conditions do not result in a multiplication in the effect in terms of bias after applying the procedure. Furthermore, deviations in $(WX)$ result in a wider spread while deviations from independence between $Z$ and $X$ result in more systematic bias. In addition, these boxplots again illustrate that in situations of an imbalance in $XZ$, applying a procedure to obtain a representative audit sample is essential. 

\subsubsection{Bias in XW}
\begin{figure}[h!]
\caption{Boxplots of the distribution of the bias of the estimated proportions of $X$ conditional on $W$, based on 1000 replicates per simulation condition. The three panels represent the three sets of bivariate relationships specified in the study, and the four columns per panel illustrate the four alternative strengths of those relationships specified (see Table 1). \label{fig:biasXW}}
\centering
\includegraphics[width=\textwidth]{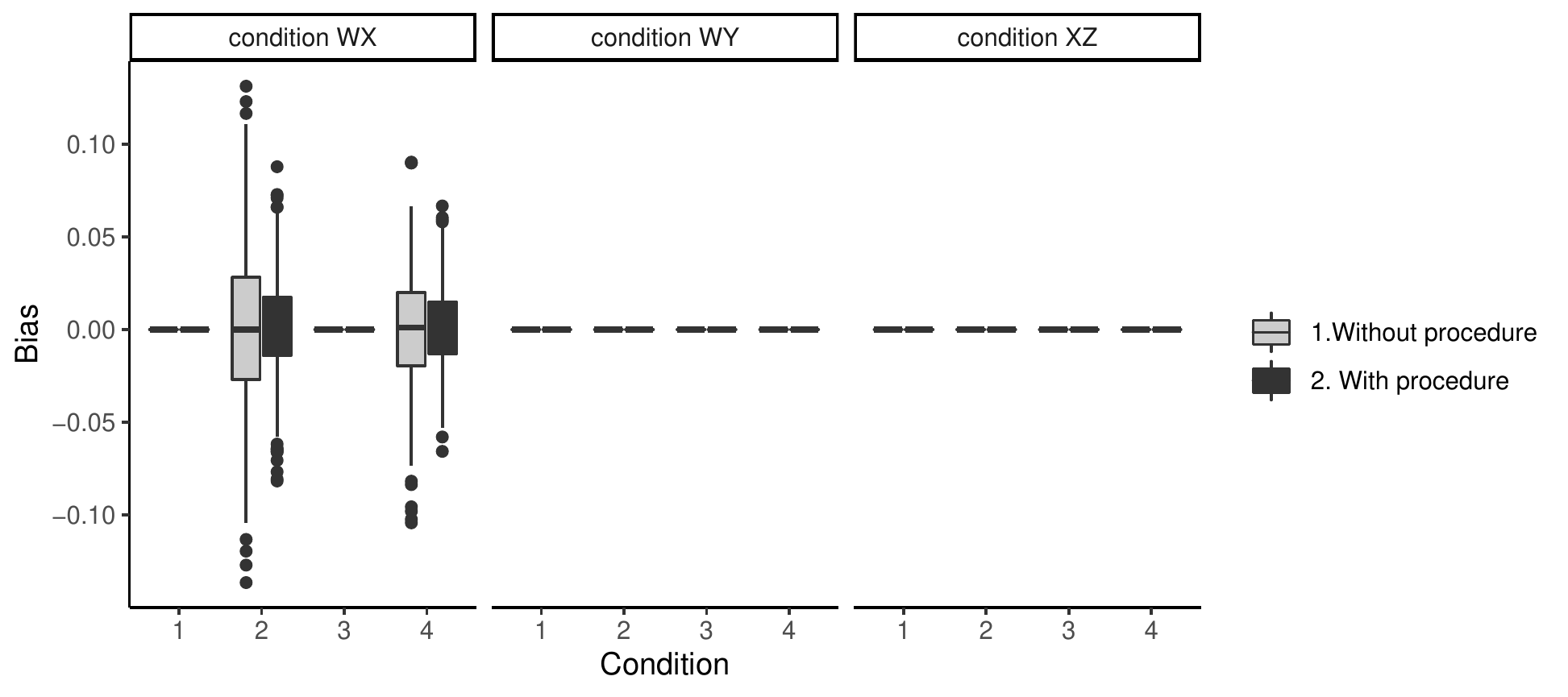}
\end{figure}

Figure \ref{fig:biasXW} illustrates the different values obtained for bias in $X \mid W$. Only the results for $(X=1 \mid W=1)$ are shown as the results for other proportions were very similar.
In the left panel, the results are shown for the four different $WX$ relationships. Similarly as when evaluating the bias for $W$, $X$ is the observed target variable of interest and $W$ its true version. It is therefore expected that no bias is present in results for $X\mid W$ if $W=X$. In conditions 2 and 4, $W\neq X$ and small increases in bias can be detected as a consequence, while in condition 3 there is also no bias present due to the fact that here $W=X$ for $X=1$. Similarly, no bias is present in $X\mid W$ under the different simulation conditions for $WY$ and $XZ$, as under these conditions $W=X$ as well. In cases of observed bias, the spread of the bias is smaller when the procedure is applied compared  to when the procedure is not applied. 

\begin{figure}[h!]
\caption{Boxplots of the distribution of the bias of the estimated proportions of $X$ conditional on $W$ before and after applying the procedure, based on 1000 replicates per simulation condition for eight combinations of different strengths of relationships between the variables $(WX)$, $(WY)$ and $(YZ)$ (see Table 1). \label{fig:biasXW_ff}}
\centering
\includegraphics[width=\textwidth]{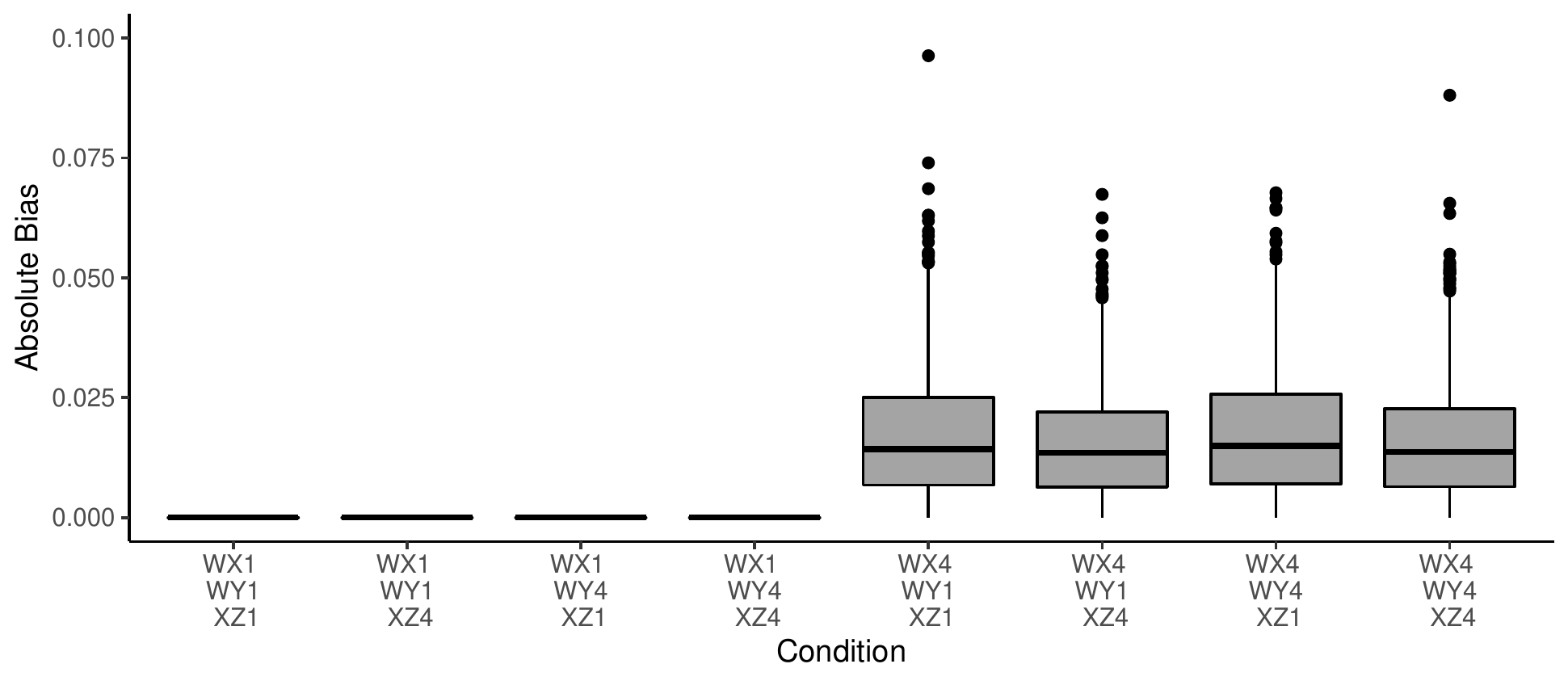}
\end{figure}

In Figure \ref{fig:biasXW_ff} it is illustrated that if $W\neq X$, generally more bias is present, although the amount of bias remains limited if the audit sampling procedure is applied. Furthermore, it is shown that combinations of measurement error in $W$ ($W\neq X$), different strengths of $WY$ and selectivity in the initial audit sample, $XZ$, do not cause for more substantive bias in $X\mid W$ than already caused by $WX$. Again, in such cases the bias is more substantive if the audit sample procedure has not been applied.

\subsubsection{Variance estimation}
Table~\ref{tab:sesd_W} and Table~\ref{tab:sesd_X|W} summarize the results of the separate simulation study mentioned in Section~\ref{sec:simvar}, to evaluate the performance of the variance estimators defined in Equation \eqref{eq:var_PW} and Equation \eqref{eq:var_PX|W} for a fixed target population. Each se-sd ratio in these tables represents the ratio of the average standard error (se) and the empirical standard deviation (sd) across 1,000 simulations, for a particular condition and target parameter $P^{W}_{w}$ (Table~\ref{tab:sesd_W}) or $P^{X|W}_{x|w}$ (Table~\ref{tab:sesd_X|W}). Ideally, this se-sd ratio would be equal to $1$.

For the estimated $P^{W}_{w}$ the empirical sd is also shown in Table~\ref{tab:sesd_W}. It is seen that the sd values were similar for all four conditions where the initial audit sample was selective. For the estimated $P^{X|W}_{x|w}$ the empirical sd values are omitted here to save space; again, these values were similar across all four conditions. 

\begin{table}[!h]
    \centering
    \begin{tabular}{lrrrcrrr}
    \hline
     & \multicolumn{3}{c}{True sd} &  & \multicolumn{3}{c}{Se-sd ratio} \\
     \cline{2-4} \cline{6-8}
    Condition & $w=1$ & $w=2$ & $w=3$ &  & $w=1$ & $w=2$ & $w=3$ \\
    \hline
    WX4,WY1,XZ4 & 0.012 & 0.013 & 0.013 &  & 1.22 & 1.23 & 1.18 \\
    WX4,WY2,XZ4 & 0.012 & 0.013 & 0.015 &  & 1.50 & 1.38 & 1.29 \\
    WX4,WY3,XZ4 & 0.012 & 0.014 & 0.015 &  & 1.20 & 1.15 & 1.13 \\
    WX4,WY4,XZ4 & 0.012 & 0.014 & 0.014 &  & 1.40 & 1.35 & 1.38 \\
    \hline
    WX4,WY1,XZ1* & 0.019 & 0.020 & 0.019 &  & 1.03 & 0.99 & 1.03 \\
    \hline
    \end{tabular}
    \caption{Estimated true standard deviation (true sd) and se-sd ratio based on 1,000 simulations for the estimated proportions of $W$. Rows refer to different simulation conditions, columns to different categories of $W$. (* = no optimization.)}
    \label{tab:sesd_W}
\end{table}

\begin{table}[!h]
    \centering
    \begin{tabular}{lrrrcrrrcrrr}
    \hline
     & \multicolumn{3}{c}{$w=1$} & & \multicolumn{3}{c}{$w=2$} & & \multicolumn{3}{c}{$w=3$} \\
     \cline{2-4} \cline{6-8} \cline{10-12}
    Condition & $x=1$ & $x=2$ & $x=3$ & & $x=1$ & $x=2$ & $x=3$ & & $x=1$ & $x=2$ & $x=3$ \\
    \hline
    WX4,WY1,XZ4 & 1.18 & 1.11 & 1.19 & & 1.20 & 1.27 & 1.27 & & 1.27 & 1.41 & 1.45 \\
    WX4,WY2,XZ4 & 1.08 & 1.07 & 1.06 & & 1.11 & 1.19 & 1.14 & & 1.28 & 1.23 & 1.35 \\
    WX4,WY3,XZ4 & 1.21 & 1.18 & 1.17 & & 1.19 & 1.27 & 1.27 & & 1.22 & 1.33 & 1.35 \\
    WX4,WY4,XZ4 & 1.13 & 1.11 & 1.09 & & 1.18 & 1.25 & 1.14 & & 1.15 & 1.18 & 1.28 \\
    \hline
    WX4,WY1,XZ1* & 0.99 & 0.98 & 0.95 & & 1.02 & 1.02 & 0.97 & & 1.00 & 1.01 & 1.02 \\
    \hline
    \end{tabular}
    \caption{Results in terms of se-sd ratio based on 1,000 simulations for the estimated proportions of $X$ conditional on $W$. Rows refer to different simulation conditions, columns to different combinations of categories of $W$ and $X$. (* = no optimization.)}
    \label{tab:sesd_X|W}
\end{table}

It is seen that for all conditions where the initial audit sample was selective, the variance estimators as defined in Equation \eqref{eq:var_PW} and Equation \eqref{eq:var_PX|W} tended to  overestimate the true variance; the overestimation in terms of standard error varied between $6\%$ and $50\%$, which in practice could be considered a moderate bias. As a benchmark, we also included a condition where $(X,Z)$ is given by the left-most matrix in Table~\ref{tab:simconds}, so that the initial audit sample already satisfied the independence model. In this case we did not apply the optimization procedure to alter the initial sample. Here it is seen that both variance estimators performed much better (last line in Tables~\ref{tab:sesd_W} and \ref{tab:sesd_X|W}). Thus, for a truly random sample from the independence model, our variance estimators are approximately correct, and the overestimation of the variance for the other conditions is due to the optimization procedure. Apparently, by minimizing the deviance, the distribution of possible audit samples obtained by this procedure is more restricted than a stratified simple random sampling design, and this is not reflected by Equation \eqref{eq:var_PW} and Equation \eqref{eq:var_PX|W}.

Within the selective conditions, it is seen that the degree of overestimation in both tables differs between, on the one hand, the first and third condition and, on the other hand, the second and fourth condition of $(W,Y)$. From Table~\ref{tab:simconds}, it is seen that this distinction corresponds to associations between $W$ and $Y$ that are relatively strong and relatively weak, respectively. For the estimated proportions $\hat{P}^{W}_{w}$, the overestimation by variance estimator \eqref{eq:var_PW} in Table~\ref{tab:sesd_W} appears to be smaller when the association between $W$ and $Y$ is stronger. Surprisingly, for the estimated error probabilities $\hat{P}^{X|W}_{x|w}$ and variance estimator \eqref{eq:var_PX|W}, the opposite effect is seen in Table~\ref{tab:sesd_X|W}.

\section{Application}

Since the introduction of the sustainable development goals \citep{cf2015transforming}, the interest in statistics regarding energy consumption has increased substantially, including many statistics which have never been produced before. One statistic of interest is the total energy consumption of establishments per economic sector. To obtain these statistics, a statistical register is required containing all companies, the economic sector they operate in and the address of their location. To construct this database, multiple incomplete administrative sources can be combined on unit level, where the unit in this case is an establishment. Many of these different data sources contain information about the economic activity of the establishment, and the observed economic activity per establishment may vary over the sources. The economic activity is  classified according to the NACE rev 2. codes \citep{RAMON2008}.

To investigate the quality of the constructed database, especially of the final code for the economic activity, an audit should be performed on a representative sample of the population of establishments. Some units (a non-representative sample) have already been audited throughout the year, and to efficiently perform an audit on a representative sample, as many of the previously audited units should be included in this new audit sample. 

The combined administrative data-set comprises $2,037,088$ establishments in $2019$. Of these units, on $173,617$ either a previous audit has been performed or the NACE codes were equal over three or more different data sources. However, the distribution of audited units is not representative for the distribution of total units with respect to background variable $Y$ and target variable $X$. Here, the target variable $X$ concerns $21$ economic sectors, which are first digit NACE codes, while background variable $Y$ contains six categories:
\begin{itemize}
    \item LV: Low voltage
    \item HV: High voltage
    \item SG: Small gas consumption
    \item MG: Middle gas consumption
    \item LG: Large gas consumption
    \item OPC: other profiles
\end{itemize}

\begin{figure}[h!]
\caption{Heatmap of the proportion of establishments per economic sector for different profile categories: for the combined administrative data set (upper panel), for the original audit sample (middle panel) and for the audit sample adjusted with the proposed procedure (lower panel). Note that we excluded the category `OPC' as it highly affected the interpretability of the heatmap.
 \label{fig:application}}
\centering
\includegraphics[width=\textwidth]{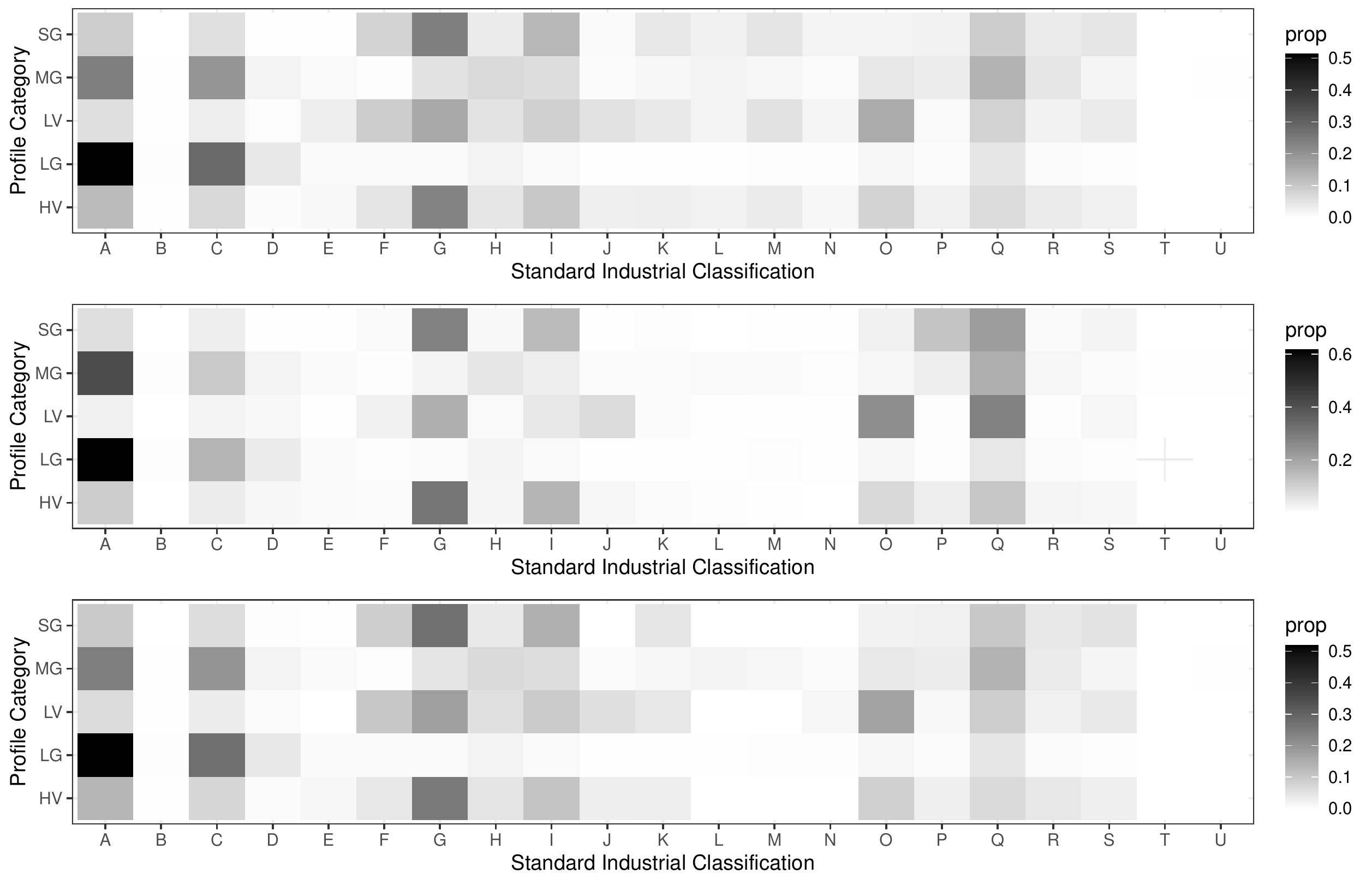}
\end{figure}

The upper heat map of Figure \ref{fig:application} illustrates the observed proportions of economic sectors $X$ per profile category $Y$. The purpose of the intended audit is to investigate the quality of these economic sector codes. In this heatmap we for example see that in the 'other profiles' category, most units are in the economic sector ``public administration and defense, compulsory social security" (``O"). In the category of large gas consumers, most units are in the agricultural sector (``A") followed by industry (``C"). In the category of small gas consumers, most units are in the sector wholesale and retail (``G"). 

Of all units shown in the upper heat map of Figure \ref{fig:application}, a subset has previously been audited, and the final economic sector codes for these units are known and can be found in the middle heat map of Figure \ref{fig:application}. By comparing the two heat maps, it can be clearly seen that given certain profiles certain sectors have been more thoroughly audited than others. For example, units in the economic sector ``human health and social work activities" (``Q") have been audited relatively thoroughly, while audits for sectors such as water and waste (``E"), construction (``F"), finance (``K") real estate (``L"), science (``M") and administration (``N") are relatively scarce. 

To obtain an audit sample that is more representative in terms of the economic sectors and with respect to the background profiles, we applied the methodology introduced in this manuscript. We first applied the methodology discussed in Section 3. Here, we applied the method multiple times using audit sample sizes $500$, $1000$, $1500$ and $2000$ ($M_{+}$), and for each sample size we varied the number of cases to exclude from the initial sample ($M_{-}$). The size of $M_{-}$ was specified as a factor with respect to $M_{+}$, and we used the values of $50$, $100$ and $120$ for this factor. Here, we investigated the resulting deviance values and concluded that we would probably be able to obtain a representative audit sample of a size between $1000$ and $1500$ and we concluded that we would need to increase the factor for removing cases. Therefore, we started with an audit sample of $1500$ and were able to reduce the sample size to $1200$. We tried to reduce the sample size even further to $1100$ but this did not result in a representative sample by using our deviance criterion. 

Now that we knew that we were able to obtain a representative sample by sampling approximately $1200$ new cases, we applied the  adjusted procedure for the objective function $F_2$ as described in Section~\ref{sec:recycle} to ensure that all additionally sampled cases were indeed not in the initial audit. To assess the deviance value, as suggested in Section~\ref{sec:procedure}, we compared it to a chi-square distribution with $J\times(I-1)$ degrees of freedom, where $I$ and $J$ are the number of categories of $X$ and $Y$, respectively, Here, with $I = 21$ and $J = 6$, we used a chi-square distribution with $120$ degrees of freedom. Choosing $\alpha = .05$, we found a value of 147 as an approximate cut-off point. Thus, if we could draw an audit sample with a deviance lower than 147, we would find this acceptable in terms of representativity.

Additionally, we tried to reduce the sample size even further, but this was not successful. Finally, we selected a sample of $1200$ and we removed $144000$ of the units from the initial audit sample (multiplication factor of $120$), with a deviance of $117.27$. 

The last heat map in Figure \ref{fig:application} illustrates the selected audit sample. This audit sample is a combination of units from the initial audit (middle heat map), units removed from this initial audit, and additionally selected units for audit ($1200$ units). In this heat map, it can be seen that certain economic sectors that were underrepresented in the initial audit are now more prominently present, such as water and waste, construction, finance and administration (``F", ``J", ``K" and ``N"). Economic sectors real estate and science (``L" and ``M") are still underrepresented. However, overall it can be concluded that the distribution of the final audit sample is more similar to the distribution in the combined administrative data set compared to the initial audit sample, while the final audit comprises a smaller selection of units. 

\section{Conclusion}

In this paper, we introduced a method that can be used to efficiently select a representative audit sample if the situation is not straightforward. For example, if a selective sample that is not representative with respect to some background characteristics has already been audited, and a representative audit sample is desired that re-uses the available sample as much as possible. 

More specifically, our method uses the joint distribution between the observed variable of interest, the background characteristics and an indicator for initial audit inclusion. Furthermore, the method assumes that there is no direct association between the true variable of interest and initial audit inclusion conditional on the observed variable of interest and the background characteristics. Here, we can analyse the previously mentioned joint distribution using an independence model with respect to the background characteristics. The fit in terms of deviance of this model can be compared to that of the saturated model. Our method then numerically searches a new solution that minimises the deviance by including new cases and excluding already audited cases. 

The simulation study illustrated that particularly when there is measurement error in the observed target variable, this can cause biased estimates of the target variable if the audit sample selection procedure is not applied to obtain a more representative audit sample. In addition, the study illustrated that without applying the procedure, bias in estimates of the target variable is obtained in situations where inclusion in the initial audit is related to the scores of the target variable of interest. This conclusion is particularly relevant, as there is in practice often a relation between the scores on the target variable and initial audit inclusion. With financial audits, the largest companies are for example often audited by default. Furthermore, we conclude from our simulation study that the variance when based on simple random sampling is often overestimated compared to the true variance, due to the fact that by minimizing the deviance, the distribution of possible audit samples is more restricted than the distribution of simple random samples would be.

The reason why we developed this procedure was that we wanted to perform an audit on establishments to investigate whether classification errors occurred in the economic sector that is used to for a publication on energy consumption per economic sector. An audit sample was already available, but this was very selective in the sense that not all economic sectors were evenly distributed. Therefore we developed, investigated and applied the method introduced in this paper. By applying this method, we were able to select an audit sample that was more representative with respect to the economic sectors, that utilized the cases that were already audited and we had control over the size of the new audit sample. 

We expect that the method is also interesting for other applications. For instance, assume that we have trained and tested a supervised machine learning model to derive a new variable. For instance, \citet{Tollenaar2018} describes that they were interested to predict whether potential crimes in police records concern cyber crime or not. Unfortunately, the annotated set that was used to train and test the model was selective, because keywords were used to seek for  cases likely to be cyber crime rather than taking a randomised sample. Consequently, the model prediction error in the test set might not be representative for the true prediction error in the population. The police records also contain background variables ($Y$) that relate to the target variable, such as crime type and whether the case has been declared by a victim or not. One could then use the predicted cyber crime in all police records by the originally trained machine learning model as a proxy ($X$) for the true cyber variable ($W$) which is available in the selective set. Next, one could apply the procedure to select additional units to generate a test set that is more more representative for the target population. The additionally selected units could then be manually annotated to obtain $W$. Finally, one could use this adjusted test set to obtain a more reliable estimate of the prediction error of the model. 

In this paper we illustrated that the method can be used in different ways. For example, an audit sample can be selected that truly has the lowest deviance, or a trade-off can be made between obtaining an acceptable value for the deviance and maximizing the amount of cases from the initial audit to be re-used. When applying the method in practice, the auditor should be aware of the practical implications when selecting a certain amount of cases to leave out or additional cases to include, particularly because including more cases in the audit will improve the probability to meet the deviance criterion on the one hand, but are also combined with increasing costs to perform the audit in practice on the other hand.

Further research should give more insight in how the method performs when different implementations of the method are chosen. In addition, more practical applications should provide insight into what further improvements are interesting for researchers. For example, are users of the method typically interested in selecting a certain audit sample of a maximum size that is within their budget? Or is this often more flexible and are users more interested in an optimal combination of smallest sample size and appropriate representativity? Finally, the results in terms of variance invite for a more thorough investigation if the audit samples drawn using this method will also be used to draw conclusions in terms of variance of estimates. 

\newpage
\printbibliography

@article{hearnshaw2003audits,
  title={Are audits wasting resources by measuring the wrong things? A survey of methods used to select audit review criteria},
  author={Hearnshaw, HM and Harker, RM and Cheater, FM and Baker, RH and Grimshaw, GM},
  journal={BMJ Quality \& Safety},
  volume={12},
  number={1},
  pages={24--28},
  year={2003},
  publisher={BMJ Publishing Group Ltd}
}

@techreport{Tollenaar2018,
  title={Predictieve textmining in politieregistraties: cyber- en gedigitaliseerde criminaliteit},
  author={Tollenaar, N. and Rokven, J. and Macro, D. and Beerthuizen, M. and van der Laan, A. M.},
  year={2018},
  publisher={Rapport van het Wetenschappelijk Onderzoek- en Documentatiecentrum, Ministerie van Justitie en Veiligheid}
}

@article{chataway2004herpes,
  title={Herpes simplex encephalitis: an audit of the use of laboratory diagnostic tests},
  author={Chataway, J and Davies, NWS and Farmer, S and Howard, RS and Thompson, EJ and Ward, KN},
  journal={Qjm},
  volume={97},
  number={6},
  pages={325--330},
  year={2004},
  publisher={Oxford University Press}
}

@article{derks2019jasp,
  title={JASP for Audit: Bayesian Tools for the Auditing Practice},
  author={Derks, Koen and de Swart, Jacques and Wagenmakers, Eric-Jan and Wille, Jan and others},
  year={2019},
  publisher={PsyArXiv}
}

@article{World2019,
  title={Listed domestic companies},
  author={TheWorldBank},
  year={2019},
  url={https://data.worldbank.org/indicator/CM.MKT.LDOM.NO}
}

@article{elder2013audit,
  title={Audit sampling research: A synthesis and implications for future research},
  author={Elder, Randal J and Akresh, Abraham D and Glover, Steven M and Higgs, Julia L and Liljegren, Jonathan},
  journal={Auditing: A Journal of Practice \& Theory},
  volume={32},
  number={sp1},
  pages={99--129},
  year={2013},
  publisher={American Accounting Assocation}
}

@article{sobel1986platonic,
  title={Platonic and Operational True Scores in Covariance Structure Analysis: An Invited Comment on {B}ielby's “{A}rbitrary Metrics in Multiple Indicator Models of Latent Variables”},
  author={Sobel, Michael E and Arminger, Gerhard},
  journal={Sociological Methods \& Research},
  volume={15},
  number={1-2},
  pages={44--58},
  year={1986},
  publisher={Sage Publications}
}

@article{scholtus2015error,
	title = {Modelling Measurement Error to Estimate Bias in Administrative and Survey Variables},
	author = {S. Scholtus and B.~F.~M. Bakker and A. van Delden},
	note = {{Discussion Paper 2015-17, Statistics Netherlands, The Hague}},
	year = {2015}
}

@book{agresti2013categorical,
  title={Categorical data analysis},
  author={Agresti, Alan},
  edition={3},
  year={2013},
  publisher={John Wiley \& Sons}
}

@book{bishopfienbergholland,
  title={Discrete Multivariate Analysis: Theory and Applications},
  author={Bishop, Yvonne M and Fienberg, Stephen E and Holland, Paul W},
  year={1975},
  publisher={MIT Press}
}

@Manual{R2019,
    title = {R: A Language and Environment for Statistical Computing},
    author = {{R Core Team}},
    organization = {R Foundation for Statistical Computing},
    address = {Vienna, Austria},
    year = {2019},
    url = {https://www.R-project.org/},
  }

@book{cochran,
  title={Sampling Techniques},
  edition={3},
  author={Cochran, W G},
  year={1977},
  publisher={John Wiley \& Sons}
}

@article{holtsmith1979,
  title={Post Stratification},
  author={Holt, D and Smith, T M F},
  journal={Journal of the Royal Statistical Society, Series A},
  volume={142},
  number={1},
  pages={33--46},
  year={1979}
}

@article{RAMON2008,
  title={RAMON - Reference And Management Of Nomenclatures},
  author={Eurostat},
  year={2008},
  url={https://ec.europa.eu/eurostat/ramon/nomenclatures/index.cfm?TargetUrl=LST_NOM_DTL&StrNom=NACE_REV2&StrLanguageCode=EN&IntPcKey=&StrLayoutCode=HIERARCHIC}
}

@article{ougrin2006clinical,
  title={Clinical audit},
  author={Ougrin, Dennis and Banarsee, Ricky},
  journal={BMJ},
  volume={333},
  number={7563},
  pages={s68--s69},
  year={2006},
  publisher={British Medical Journal Publishing Group}
}

@article{bell1995auditing,
  title={Auditing practice, research, and education: A productive collaboration},
  author={Bell, Timothy B and Bell, Arnold M and {American Accounting Association} and others},
  year={1995},
  publisher={American Institute of Certified Public Accountants}
}

@inproceedings{zadrozny2004learning,
  title={Learning and evaluating classifiers under sample selection bias},
  author={Zadrozny, Bianca},
  booktitle={Proceedings of the twenty-first international conference on Machine learning},
  pages={114},
  year={2004}
}

@article{klingwort2021using,
  title={Inferring network traffic from sensors without a sampling design},
  author={Klingwort, Jonas and Burger, Joep and Buelens, Bart},
  year={2021},
  publisher={CBS Discussion Paper}
}

@article{rubin1976inference,
  title={Inference and missing data},
  author={Rubin, Donald B},
  journal={Biometrika},
  volume={63},
  number={3},
  pages={581--592},
  year={1976},
  publisher={Oxford University Press}
}

@article{cf2015transforming,
  title={Transforming our world: the 2030 Agenda for Sustainable Development},
  author={{UN General Assembly}},
  year={2015},
  url={https://www.refworld.org/docid/57b6e3e44.html}
}
\end{document}